\documentstyle[aps,epsfig]{revtex}

\newcommand{\shat}{\hat{s}}
\newcommand{\gev}{{\rm GeV}}

\newcommand{\gw}{\Gamma_W}

\newcommand{\pbar}{\overline{p}}
\newcommand{\invpb}{{\rm pb}^{-1}}
\newcommand{\wlnu}{W\rightarrow \ell\nu}
\newcommand{\wenu}{W\rightarrow e\nu}
\newcommand{\wmunu}{W\rightarrow \mu\nu}
\newcommand{\met}{{\not\!\!E}_T}
\begin{document}
\draft
\title{
\begin{flushright}                                           
\rm CDF/PUB/ELECTROWEAK/PUBLIC/5148 \\
\rm Fermilab-Pub-00/085-E \\
\rm V1.6 2000-04-15 \\
\end{flushright}                                             
\boldmath
Direct Measurement of the $W$ Boson Width in $p\overline{p}$
Collisions at $\sqrt{s}=1.8~{\rm TeV}$
}
\author{}
\address{}
\date{\today}
\maketitle
\font\eightit=cmti8
\def\r#1{\ignorespaces $^{#1}$}
\hfilneg
\begin{sloppypar}
\noindent
T.~Affolder,\r {21} H.~Akimoto,\r {43}
A.~Akopian,\r {36} M.~G.~Albrow,\r {10} P.~Amaral,\r 7 S.~R.~Amendolia,\r {32} 
D.~Amidei,\r {24} K.~Anikeev,\r {22} J.~Antos,\r 1 
G.~Apollinari,\r {10}
% ...
\end{sloppypar}
\vskip .026in
\begin{center}
(CDF Collaboration)
\end{center}

\vskip .026in
\begin{center}
\r 1  {\eightit Institute of Physics, Academia Sinica, Taipei, Taiwan 11529, 
Republic of China} \\
\r 2  {\eightit Argonne National Laboratory, Argonne, Illinois 60439} \\
\r 3  {\eightit Istituto Nazionale di Fisica Nucleare, University of Bologna,
I-40127 Bologna, Italy} \\
\r 4  {\eightit Brandeis University, Waltham, Massachusetts 02254} \\
\r 5  {\eightit University of California at Los Angeles, Los 
Angeles, California  90024} \\  
\r 6  {\eightit Instituto de Fisica de Cantabria, CSIC-University of Cantabria, 
39005 Santander, Spain} \\
\r 7  {\eightit Enrico Fermi Institute, University of Chicago, Chicago, 
Illinois 60637} \\
\r 8  {\eightit Joint Institute for Nuclear Research, RU-141980 Dubna, Russia}
\\
\r 9  {\eightit Duke University, Durham, North Carolina  27708} \\
\r {10}  {\eightit Fermi National Accelerator Laboratory, Batavia, Illinois 
60510} \\
\r {11} {\eightit University of Florida, Gainesville, Florida  32611} \\
\r {12} {\eightit Laboratori Nazionali di Frascati, Istituto Nazionale di Fisica
               Nucleare, I-00044 Frascati, Italy} \\
\r {13} {\eightit University of Geneva, CH-1211 Geneva 4, Switzerland} \\
\r {14} {\eightit Harvard University, Cambridge, Massachusetts 02138} \\
\r {15} {\eightit Hiroshima University, Higashi-Hiroshima 724, Japan} \\
\r {16} {\eightit University of Illinois, Urbana, Illinois 61801} \\
\r {17} {\eightit The Johns Hopkins University, Baltimore, Maryland 21218} \\
\r {18} {\eightit Institut f\"{u}r Experimentelle Kernphysik, 
Universit\"{a}t Karlsruhe, 76128 Karlsruhe, Germany} \\
\r {19} {\eightit Korean Hadron Collider Laboratory: Kyungpook National
University, Taegu 702-701; Seoul National University, Seoul 151-742; and
SungKyunKwan University, Suwon 440-746; Korea} \\
\r {20} {\eightit High Energy Accelerator Research Organization (KEK), Tsukuba, 
Ibaraki 305, Japan} \\
\r {21} {\eightit Ernest Orlando Lawrence Berkeley National Laboratory, 
Berkeley, California 94720} \\
\r {22} {\eightit Massachusetts Institute of Technology, Cambridge,
Massachusetts  02139} \\   
\r {23} {\eightit Institute of Particle Physics: McGill University, Montreal 
H3A 2T8; and University of Toronto, Toronto M5S 1A7; Canada} \\
\r {24} {\eightit University of Michigan, Ann Arbor, Michigan 48109} \\
\r {25} {\eightit Michigan State University, East Lansing, Michigan  48824} \\
\r {26} {\eightit University of New Mexico, Albuquerque, New Mexico 87131} \\
\r {27} {\eightit The Ohio State University, Columbus, Ohio  43210} \\
\r {28} {\eightit Osaka City University, Osaka 588, Japan} \\
\r {29} {\eightit University of Oxford, Oxford OX1 3RH, United Kingdom} \\
\r {30} {\eightit Universita di Padova, Istituto Nazionale di Fisica 
          Nucleare, Sezione di Padova, I-35131 Padova, Italy} \\
\r {31} {\eightit University of Pennsylvania, Philadelphia, 
        Pennsylvania 19104} \\   
\r {32} {\eightit Istituto Nazionale di Fisica Nucleare, University and Scuola
               Normale Superiore of Pisa, I-56100 Pisa, Italy} \\
\r {33} {\eightit University of Pittsburgh, Pittsburgh, Pennsylvania 15260} \\
\r {34} {\eightit Purdue University, West Lafayette, Indiana 47907} \\
\r {35} {\eightit University of Rochester, Rochester, New York 14627} \\
\r {36} {\eightit Rockefeller University, New York, New York 10021} \\
\r {37} {\eightit Rutgers University, Piscataway, New Jersey 08855} \\
\r {38} {\eightit Texas A\&M University, College Station, Texas 77843} \\
\r {39} {\eightit Texas Tech University, Lubbock, Texas 79409} \\
\r {40} {\eightit Istituto Nazionale di Fisica Nucleare, University of Trieste/
Udine, Italy} \\
\r {41} {\eightit University of Tsukuba, Tsukuba, Ibaraki 305, Japan} \\
\r {42} {\eightit Tufts University, Medford, Massachusetts 02155} \\
\r {43} {\eightit Waseda University, Tokyo 169, Japan} \\
\r {44} {\eightit University of Wisconsin, Madison, Wisconsin 53706} \\
\r {45} {\eightit Yale University, New Haven, Connecticut 06520} \\
\end{center}

\begin{abstract}
\begin{center}
{\bf Abstract}
\end{center}
This Letter describes a direct measurement of the $W$ boson total
decay width, $\gw$, using the Collider Detector at Fermilab.  The
measurement uses an integrated luminosity of $90~\invpb$, collected
during the 1994--1995 run of the Fermilab Tevatron $p\pbar$ collider.
The width is determined by normalizing predicted signal and background
distributions to 49844 $\wenu$ candidates and 21806 $\wmunu$
candidates in the transverse-mass region $M_T<200~\gev$ and then
fitting the predicted shape to the 438 electron events and 196 muon
events in the high-$M_T$ region, $100<M_T<200~\gev$.  The result is
$\Gamma_W=2.04\pm0.11$~(stat)~$\pm0.09$~(syst)~GeV.
\end{abstract}

\pacs{13.38.Be}

The masses and coupling strengths of the gauge bosons that mediate the
known forces are fundamental parameters in the Standard Model (SM).
The $W$ boson width, $\gw$, is precisely predicted in terms of these
masses and couplings.  The leptonic partial width $\Gamma(\wlnu)$ for
the lepton $\ell$ can be expressed as $G_F M_W^3 / 6\sqrt{2}\pi
(1+\delta_{\rm SM})$ in terms of the well-measured muon decay constant
$G_F$, the $W$ boson mass $M_W$, and a small ($<\frac{1}{2}\%$)
radiative correction $\delta_{\rm SM}$ to the Born-level
expression~\cite{Rosner}.  Dividing the partial width by the branching
ratio, $B(\wlnu)=1/(3+6(1+\alpha_s(M_W)/\pi+{\cal O}(\alpha_s^2)))$,
gives the SM prediction for the full width of the $W$ boson,
$\gw=2.093\pm0.002~\gev$~\cite{PDG}.

The $W$ width has been measured indirectly using the ratio $R \equiv
{\sigma B(p\overline{p}\rightarrow W \rightarrow \ell\nu) \over \sigma
B(p\overline{p}\rightarrow Z^{0} \rightarrow
\ell^{+}\ell^{-})}$~\cite{Cabibbo}, with a current precision of
50~MeV~\cite{RunIb-R}, by assuming SM values for $\sigma(W)/\sigma(Z)$
and $\Gamma(W\rightarrow \ell\nu)$ and using the LEP measurement of
the branching ratio $B(Z\rightarrow \ell^{+}\ell^{-})$.  Direct
measurements of $\Gamma_W$ from lineshape analyses have been reported
in $p\overline{p}$ collisions with a precision of
320~MeV~\cite{CDF-direct} and in $e^{+}e^{-}$ collisions where
presently the most precise measurement has an uncertainty of
375~MeV~\cite{LEP-direct}.

The CDF collaboration previously reported~\cite{CDF-direct} a direct
measurement of the $W$ width using an integrated luminosity of
20~pb$^{-1}$ of $W\rightarrow e\nu$ data collected by CDF during the
1992-1993 run of the Fermilab Tevatron collider.  This Letter extends
that measurement, using a 90~pb$^{-1}$ sample of $W\rightarrow e\nu$
and $W\rightarrow \mu\nu$ data collected by CDF during the period from
January 1994 to July 1995.

This paper presents a measurement of $\Gamma_W$ obtained in studies of
the transverse-mass spectra of leptonic $W$ decays.  The transverse
mass is defined as $M_T \equiv \sqrt{2P_T^\ell
P_T^\nu[1-\cos(\Delta\phi)]}$, where $\ell=e$ or $\mu$, $P_T^\ell$ and
$P_T^\nu$ are the transverse momenta~\cite{CDF-coord} of the charged
lepton and neutrino, and $\Delta\phi$ is the azimuthal angle between
them.  The transverse-mass spectrum exhibits a Jacobian edge at the
$W$ mass.  Events with $M_T>M_W$ arise due to a combination of the
non-zero $W$ width and the detector resolution.  A precise $\Gamma_W$
measurement from the high-mass tail is possible, however, because the
width component of the high-$M_T$ lineshape falls off much more slowly
than the resolution component.  In this analysis the $W$ width is
determined from a binned log-likelihood fit to the transverse-mass
distribution in the region $100<M_T<200$~GeV.  The choice $M_T>100$~GeV
minimizes the sum in quadrature of systematic and statistical
uncertainties.

The portions of the CDF detector relevant to this analysis are
described briefly below.  Detailed descriptions can be found
elsewhere~\cite{detector}.  Electromagnetic and hadronic calorimeters,
arranged in a projective tower geometry, cover the pseudorapidity
range $|\eta|<4.2$.  In the region $|\eta|<1.0$, a lead/scintillator
electromagnetic calorimeter (CEM) measures electron energies with
resolution $\sigma(E)/E = {13.5\% / \sqrt{E_T~({\rm GeV})}} \oplus
1.5\%$.  A cylindrical drift chamber (CTC), immersed in a 1.4~T
solenoidal magnetic field, tracks charged particles in the range
$|\eta|<1.0$ with vertex-constrained momentum resolution
$\sigma(p_T)/p_T = 0.09\%\times p_T~({\rm GeV})$.  A system of drift
chambers and steel absorber identifies muons in the region
$|\eta|<1.0$.  Finally, a time-projection chamber (VTX) finds
$p\overline{p}$ interaction vertices along the $z$ axis.

Candidate $W\rightarrow e\nu$ events are required to have an electron
in the central barrel region of the detector ($|\eta|<1.0$) with CEM
transverse energy $E_T^e>25$~GeV and CTC transverse momentum
$p_T^e>15$~GeV.  The electron track must be isolated in the CTC,
having no other track with $p_T>1$~GeV within a cone in the
$\eta$-$\phi$ space of $\sqrt{(\Delta\phi)^2+(\Delta\eta)^2}=0.25$
centered on the electron.  The ratio of energy in the hadron (Had) and
electromagnetic (CEM) calorimeter towers of the electron cluster is
required to satisfy $E_{\rm Had}/E_{\rm CEM}<0.055+0.00045E~({\rm
GeV})$.  The electron shower must be contained within a fiducial
region of the CEM, away from calorimeter cell boundaries, and have a
profile consistent with test-beam data.

Candidate $W\rightarrow\mu\nu$ events must have a CTC track with
transverse momentum $p_T^\mu>25$~GeV.  The CTC track must be well
matched to a track segment in the muon chambers.  The signal in the
electromagnetic and hadronic calorimeters must be consistent with the
passage of a minimum-ionizing particle, satisfying $E_{\rm CEM}<2$~GeV
and $E_{\rm Had}<6$~GeV.  Trigger prescale factors, trigger
efficiency, and limited azimuthal coverage reduce the
$W\rightarrow\mu\nu$ acceptance by a factor $\sim 2$ with respect to
the $W\rightarrow e\nu$ acceptance~\cite{mwprd}.

In both $W\rightarrow e\nu$ and $W\rightarrow\mu\nu$ candidate events,
a transverse momentum imbalance is required to signal the presence of
the neutrino.  The missing transverse energy, $\vec\met\equiv
-(\vec{P_T^\ell}+\vec{u})$, must satisfy $\met>$25~GeV, where
$P_T^\ell$ denotes $E_T^e$ for electrons or $p_T^\mu$ for muons.  The
recoil transverse energy vector, $\vec{u}$, is defined as $\sum_i E_i
\sin\theta_i (\cos\phi_i, \sin\phi_i)$, for calorimeter towers $i$
with $|\eta|<3.6$, excluding those traversed by the charged lepton.
The vector $-\vec{u}$, which includes initial state QCD radiation,
underlying event energy, and the products of overlapping
$p\overline{p}$ interactions, is an estimator of the transverse
momentum of the $W$.  To reduce backgrounds and improve transverse
mass resolution, the recoil energy must satisfy $u<20$~GeV.  To ensure
good measurements in the drift chamber and calorimeters, both
electrons and muons must pass through all 84 layers of the CTC and
must originate from an event vertex located within 60~cm of the
detector center along the $z$ axis.  Events consistent with cosmic
rays or $Z\rightarrow \ell^{+}\ell^{-}$ decays are removed.  The
$W\rightarrow e\nu$ sample consists of 49844 events in the range
$40<M_T<200$~GeV; the $W\rightarrow\mu\nu$ sample consists of 21806
events with transverse masses in the same range.

Several background processes can mimic the $W$ signal.  The process
$W\rightarrow\tau\nu\rightarrow \ell\nu\nu\nu$ has a signature similar
to $W\rightarrow \ell\nu$ decays but at lower $M_T$.  The process
$Z\rightarrow ee$, where one electron is detected and the other is
lost because it falls into an uninstrumented region of the detector,
can produce the signature of an electron and $\met$; similarly, a
$Z\rightarrow\mu\mu$ event can pass the $W\rightarrow\mu\nu$ selection
if one muon is outside the $|\eta|$ acceptance of the CTC.  Multijet
backgrounds from QCD processes arise when one jet fragments into a
single particle that mimics a charged lepton and another is
mismeasured to produce an energy imbalance.  A Monte Carlo simulation
normalized to the $W\rightarrow \ell\nu$ data is used to predict the
$W\rightarrow\tau\nu$ and $Z\rightarrow \ell^{+}\ell^{-}$ backgrounds.
The QCD contribution is estimated from a study of non-isolated leptons
in the data.  Table~\ref{table:bg} summarizes the background
contributions for the $W\rightarrow e\nu$ and $W\rightarrow\mu\nu$
samples.  In the $M_T>200~\gev$ region, 23 $W\rightarrow l\nu$
candidates are observed, consistent with the expectation of $20\pm5$
events.

Since the $W$ and $Z$ bosons share a common production mechanism and
are close in mass, $Z\rightarrow \ell^{+}\ell^{-}$ decays are used
extensively to model the detector's response to $W\rightarrow \ell\nu$
events.  Samples of $Z\rightarrow ee$ and $Z\rightarrow \mu\mu$
candidates are selected using the same charged-lepton requirements as
for $W\rightarrow \ell\nu$ candidates, with the exception that one
muon from each $Z\rightarrow\mu\mu$ pair is subjected to less
stringent fiducial requirements.  The invariant mass must fall in the
window $70<M^{\ell\ell}<110$~GeV and the boson transverse momentum
must satisfy $P_T^{Z}<50$~GeV.  There are 2012 $Z\rightarrow ee$ and
1830 $Z\rightarrow \mu\mu$ candidates, with negligible background.
Using the LEP values of the $Z$-boson mass and width~\cite{PDG}, the
scales and resolutions of the lepton energy and momentum measurements
are extracted from a fit to the $Z$-candidate $M^{\ell\ell}$ spectra.
Additional fits to the $Z$ data constrain the boson
transverse-momentum spectra and provide an empirical model of the
recoil response $\vec{u}$ as a function of $P_T^{\ell\ell}$.  Details
of the recoil model can be found in Refs.~\cite{mwprd,theses}.

The $W$ transverse mass spectrum is modeled using a Monte Carlo
simulation that generates lowest-order diagrams of $W$ production,
$q\overline{q}\rightarrow W$, according to an energy-dependent
Breit-Wigner distribution:
\[
\sigma(\shat) \sim \frac{\shat}{(\shat-M_W^2)^2+\shat^2\Gamma_W^2/M_W^2},
\]
where $\sqrt{\shat}$ is the (generally off-shell) $l\nu$ mass.
The MRS-R2~\cite{pdf} parton distribution
functions are used.  The effects of higher order QCD diagrams for $W$
production are included by giving the $W$ bosons transverse momenta
according to a fit to the boson momentum spectra in the $Z\rightarrow
\ell^{+}\ell^{-}$ samples; a theoretical calculation~\cite{wpt} allows
the $W$ transverse momentum spectrum to be derived from the $Z$
transverse momentum spectrum.  The generator includes the effect of
$W\rightarrow \ell\nu\gamma$ decays, and the effect of photon
radiation on the lepton selection is accounted for in a detailed
simulation.  The lepton momenta are passed through a simulation of
the detector response, which includes a parametric model of the
$\vec{u}$ measurement as a function of boson transverse momentum.  The
same kinematic and geometric cuts as in the data are applied in the
simulation.

\begin{figure}
\epsfig{file=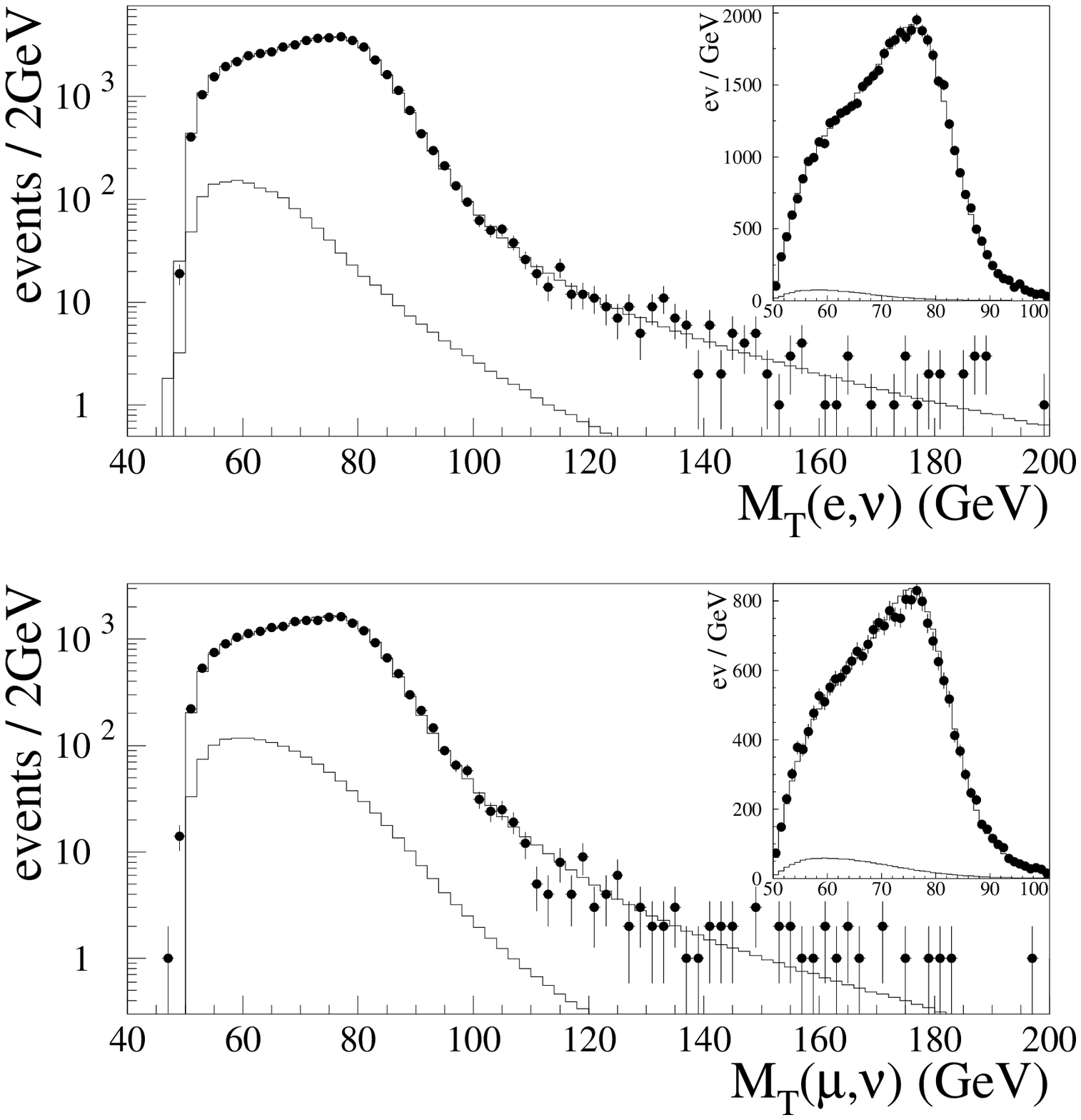,width=\columnwidth}
\caption{Transverse mass spectra (filled circles) for $W\rightarrow
e\nu$ (upper) and $W\rightarrow \mu\nu$ (lower) data, with best fits
superimposed as a solid curve.  The lower curve in each graph shows
the sum of estimated backgrounds.  Each inset shows the 50--100~GeV
region on a linear scale.}
\label{figure:mtplot}
\end{figure}

The simulation produces $M_T$ spectra for a range of $\Gamma_W$
values, from 1.0 to 3.0~GeV in 50~MeV intervals.  Each spectrum is
normalized to the number of expected signal events in the region
$M_T<200$~GeV, and background shapes at the rates shown in
Table~\ref{table:bg} are added to the predicted spectra.  A binned
likelihood fit in 1~GeV bins over the region $100<M_T<200$~GeV returns
$\Gamma_W = 2.175\pm0.125$~(stat)~GeV for the electron channel and
$\Gamma_W = 1.780\pm0.195$~(stat)~GeV for the muon channel.
Figure~\ref{figure:mtplot} shows the data with the best fits and
normalized background shapes superimposed.

The systematic uncertainties in this measurement of the $W$ width are
due to effects that alter the shape of the transverse mass
distribution.  The most important sources of uncertainty are
non-linearity of the CEM $E_T$ measurement (relevant only for
electrons), recoil modeling, the $W$ transverse momentum spectrum, and
backgrounds.  To estimate the uncertainties due to these effects,
these parameters are varied in the simulation and the simulated
transverse mass spectra with the varied input parameters are fit to
the nominal templates.

Linearity of the CEM energy measurement is studied by comparing CEM
energies to CTC momenta over the range of energies spanned by
electrons from the $W$ and $Z$ data samples.  A fit to the form
$E^{\rm meas}/E^{\rm true} = 1+\epsilon(E_T^{\rm
meas}-\left<E_T\right>)$ yields
$\epsilon=(2.9\pm1.3)\times10^{-4}~{\rm GeV}^{-1}$; a study of
$\psi\rightarrow ee$ and $\Upsilon\rightarrow ee$ events yields a
consistent value of $\epsilon$.  CEM energies in the data are
corrected event-by-event for this effect.  Taking the uncertainty on
$\epsilon$ to be $\pm2.9\times10^{-4}~\gev^{-1}$ shifts $\Gamma_W$ in
the simulation by $\mp60$~MeV.  A study of $\psi\rightarrow \mu\mu$,
$\Upsilon\rightarrow \mu\mu$, and $Z\rightarrow \mu\mu$ resonances
finds no evidence for non-linearity in the momentum measurement.
Varying the linearity of the momentum measurement within the bounds
allowed by the data changes the muon-channel $W$ width by only 5~MeV
in the simulation.

The parameters of the recoil model are varied according to the
covariance matrices obtained in the fits of $\vec{u}$ as a function of
$P_T^{\ell\ell}$ in the $Z$ data.  Because the $e$ and $\mu$ analyses
use independent fits to their respective $Z\rightarrow
\ell^{+}\ell^{-}$ samples, the uncertainties are different for the two
channels and are statistically independent.  The effect on $\Gamma_W$
is 60~MeV in the electron channel and 90~MeV in the muon channel.
Similarly, the statistical uncertainty in the fits to the transverse
momentum spectra of the two $Z$ boson samples yields a $\Gamma_W$
error of 55~MeV in the electron channel and 70~MeV in the muon
channel.  Varying the background predictions within the errors quoted
in Table~\ref{table:bg} changes the electron result by 30~MeV and the
muon result by 50~MeV.

Varying the muon identification cuts in the data and the muon
acceptance model in the simulation yield a combined $\Gamma_W$ error
of 40~MeV.  To check the detector simulation used in the electron
analysis, a sample generated using an independent simulation program
is fit with the standard $\Gamma_W$ templates and found to agree;
the statistical precision of the the check, 30~MeV, is taken as a
systematic uncertainty.

Fits to the $Z\rightarrow \ell^{+}\ell^{-}$ mass spectra determine
both the CEM energy and CTC momentum scales to 0.1\%.  Varying these
scales by 0.1\% in the simulation changes $\Gamma_W$ by 20~MeV and
15~MeV respectively in the electron and muon analyses.  Varying the
CEM and CTC resolutions within the uncertainties allowed by fits to
the $Z$ mass spectra varies $\Gamma_W$ by 10~MeV and 20~MeV
respectively.

Monte Carlo spectra have been generated using a variety of modern
parton distribution functions, including a set whose $d/u$ ratio was
modified~\cite{UnKi} to span the range allowed by CDF measurements of
the rapidity asymmetry in $W\rightarrow \ell\nu$ decay.  The RMS shift
in $\Gamma_W$ is 15~MeV in both channels.  Varying the $W$ mass by the
current world average uncertainty of 40~MeV~\cite{RunIb-R,wmass} from
the central value 80.40~GeV changes $\Gamma_W$ by 10~MeV in each
channel.  Finally a study comparing $W\rightarrow \ell\nu\gamma$ and
$W\rightarrow \ell\nu\gamma\gamma$ in the PHOTOS simulation yields a
systematic uncertainty of 10~MeV.  These three final sources of
uncertainty are common to both analyses.

Uncertainties have been calculated separately for the fit regions
$M_T>90$~GeV, $M_T>100$~GeV, and $M_T>110$~GeV.  While the statistical
uncertainty decreases as the cut value is lowered, the systematic
uncertainty increases.  The $M_T>100$~GeV fit minimizes the total
uncertainty.  The results of the $M_T>100$~GeV and $M_T>110$~GeV fits
differ by 10~MeV in the electron channel and 60~MeV in the muon
channel.

Table~\ref{table:err} summarizes the sources of uncertainty described
above.  Combining the $e$ and $\mu$ results, with a common error of
25~MeV, yields $\Gamma_W=2.04\pm0.11$~(stat)~$\pm0.09$~(syst)~GeV.
Including the previously published CDF electron
result~\cite{CDF-direct} with the same common error yields
$\Gamma_W=2.05\pm0.10$~(stat)~$\pm0.08$~(syst)~GeV.  The result is in
good agreement with the Standard Model value.

The vital contributions of the Fermilab staff and the technical staffs
of the participating institutions are gratefully acknowledged.  This
work is supported by the U.S. Department of Energy, the National
Science Foundation, the Natural Sciences and Engineering Research
Council of Canada, the Instituto Nazionale di Fisica Nucleare of
Italy, the Ministry of Education, Science and Culture of Japan, the
National Science Council of the Republic of China, and the A.P. Sloan
Foundation.

%

\begin{table}
\caption{The numbers of events in the $W\rightarrow \ell\nu$ signal
samples and the estimated numbers of background events.}
\label{table:bg}
\begin{tabular}{|l|c|c|c|c|}
 Channel
 & \multicolumn{2}{c|}{$W\rightarrow e\nu$} &
 \multicolumn{2}{c|}{$W\rightarrow\mu\nu$} \\ \hline
 $M_T$ region
 & $40$--$200$ & $100$--$200$
 & $40$--$200$ & $100$--$200$ \\ 
\hline
Events                & 49844         & 438      & 21806       & 196     \\
\hline
$W\rightarrow\tau\nu$ & $870\pm100$   & $4\pm2$  & $440\pm20$  & $2\pm2$ \\
$Z\rightarrow\ell\ell$ & $170\pm85$    & $5\pm3$  & $760\pm30$  & $5\pm2$ \\
QCD multijets         & $450\pm110$   & $10\pm4$ & $175\pm15$  & $4\pm3$ \\
cosmic rays           & $0$           & 0        & $4\pm2$     & 0       \\ 
\hline
Backgrounds           & $1490\pm170$  & $19\pm5$ & $1379\pm40$ & $11\pm4$ \\
\end{tabular}
\end{table}

\begin{table}
\caption{The sources of uncertainty on $\Gamma_W$ for the
$W\rightarrow e\nu$ and $W\rightarrow\mu\nu$ measurements.  The last
three uncertainties are common to the electron and muon analyses.}
\label{table:err}
\begin{tabular}{|l|c|c|}
 Source & $\Delta\Gamma$ ($e$,MeV) & 
          $\Delta\Gamma$ ($\mu$,MeV) \\ \hline
 Statistics & 125 & 195 \\ \hline
 Lepton $E$ or $p_T$ non-linearity & 60 & 5 \\
 Recoil model & 60 & 90 \\
 $W$ $P_T$ & 55 & 70 \\
 Backgrounds & 30 & 50 \\
 Detector modeling, lepton ID & 30 & 40 \\
 Lepton $E$ or $p_T$ scale & 20 & 15 \\
 Lepton resolution & 10 & 20 \\
 PDFs (common) & 15 & 15 \\
 $M_W$ (common) & 10 & 10 \\
 QED (common) & 10 & 10 \\ \hline
 Total systematic & 115 & 135 \\ \hline
 Total stat + syst & 170 & 235 \\
\end{tabular}
\end{table}


\begin{references}
\bibitem{CDF-coord}
CDF uses a cylindrical coordinate system in which $\phi$ is the
azimuthal angle, $\theta$ is the polar angle, and $z$ points in the
proton beam direction and is zero at the center of the detector.  The
pseudorapidity $\eta$ is defined as $\eta \equiv
\tanh^{-1}(\cos\theta)$.  The transverse plane is the plane
perpendicular to the $z$ axis.  The transverse energy is defined to be
$E_T = E\sin\theta$.
\end{references}
\end{document}